\documentclass[reprint, two column, prl, superscriptaddress, showkeys]{revtex4-1}
\usepackage{graphicx,epsfig}
\usepackage{amssymb}
\usepackage{amsmath}
\usepackage{bm}
\usepackage{textcomp}
\usepackage{color}

\begin{document}
\setcounter{page}{1}

\title[]{Observation of fractional quantum Hall effect in an InAs quantum well }
\author{Meng K. \surname{Ma}}
\author{Md. Shafayat \surname{Hossain}}
\author{K. A. \surname{Villegas Rosales}}
\author{H. \surname{Deng}}
\affiliation{Department of Electrical Engineering, Princeton University, Princeton, New Jersey 08544, USA}
\author{T. \surname{Tschirky}}
\author{W. \surname{Wegscheider}}
\affiliation{Laboratory for Solid State Physics, ETH Z{\"u}rich, 8093 Z{\"u}rich, Switzerland}
\author{M. \surname{Shayegan}}
\affiliation{Department of Electrical Engineering, Princeton University, Princeton, New Jersey 08544, USA}
\date{\today}

\begin{abstract}
The two-dimensional electron system in an InAs quantum well has emerged as a prime candidate for hosting exotic quasi-particles with non-Abelian statistics such as Majorana fermions and parafermions. To attain its full promise, however, the electron system has to be clean enough to exhibit electron-electron interaction phenomena. Here we report the observation of fractional quantum Hall effect in a very low disorder InAs quantum well with a well-width of 24 nm, containing a two-dimensional electron system with a density $n=7.8 \times 10^{11}$ cm$^{-2}$ and low-temperature mobility $1.8 \times 10^6$ cm$^2$/Vs. At a temperature of $\simeq35$ mK and $B\simeq24$ T, we observe a deep minimum in the longitudinal resistance, accompanied by a nearly quantized Hall plateau at Landau level filling factor $\nu=4/3$.  
\end{abstract} 

\maketitle   

\begin{figure*}[t!]
  \begin{center}
    \psfig{file=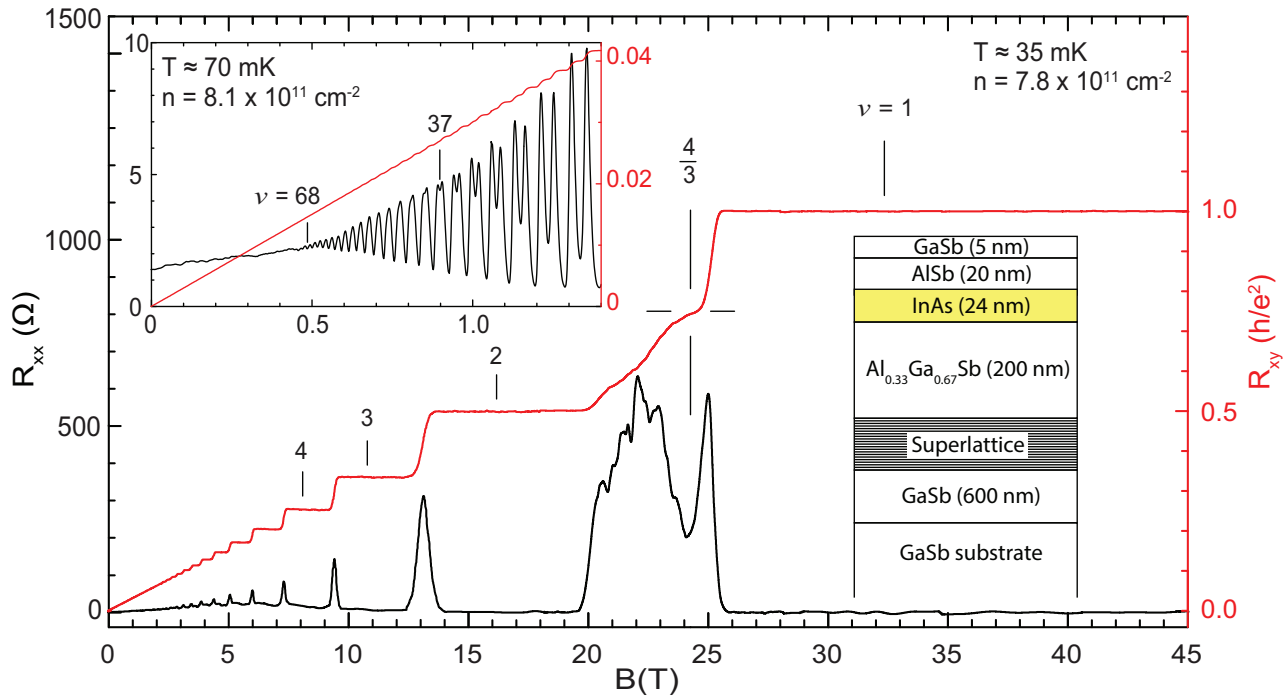, width=0.95\textwidth}
  \end{center}
  \caption{\label{fig1} The longitudinal resistance $R_{xx}$ and Hall resistance $R_{xy}$ vs. perpendicular magnetic field $B$ from 0 to 45 T for a 24-nm-wide InAs/Al$_{0.33}$Ga$_{0.67}$Sb quantum well at $\simeq35$ mK. The details of the structure are shown in the right inset. The vertical marks in the main figure indicate the field positions of integer fillings $\nu=1$ to $4$ as well as the expected $\nu=4/3$ $R_{xx}$ minimum position. The expected $R_{xy}$ quantization value ($3h/4e^2$) is marked with horizontal lines. The upper left inset shows the low-field $R_{xx}$ and $R_{xy}$. The Shubnikov-de Haas oscillations are resolvable at magnetic fields as low as $\simeq0.5$ T ($\nu=68$), and become spin-resolved for $B>0.9$ T ($\nu\leqslant37$).}
\label{fig:fullfield}
\end{figure*}

Two-dimensional electron systems (2DESs) confined to InAs quantum wells have been of interest for decades \cite{Tsui.PRL.1970,Chang.SC.1980,Yeh.JMC.2013,Brandstetter.APL.2016,Datta.APL.1990,Nitta.PRL.1997,fnote.1,Kjaergaard.PRA.2017,fnote.126,Das.Nat.2012,Finck.PRL.2013,Shojaei.APL.2015,Shabani.PRB.2016,Tschirky.PRB.2017,Hatke.arXiv.2017}. The small electron effective mass, combined with the relatively high purity of the epitaxially grown InAs layers, leads to very high electron mobilities, rendering the InAs 2DES a system of choice for high-speed transistors and sensors \cite{Yeh.JMC.2013,Brandstetter.APL.2016}. Moreover, thanks to their strong spin-orbit coupling, InAs 2DESs have long been prime candidates for spintronic devices \cite{Datta.APL.1990, Nitta.PRL.1997,fnote.1}. Interest in InAs 2DESs has surged to new heights recently because of the possibility of inducing superconductivity through the proximity effect in these systems \cite{Shabani.PRB.2016,Kjaergaard.PRA.2017}. Theoretical proposals indeed suggest that such systems might host exotic topological states of matter with non-Abelian quasi-particles, namely Majorana fermions, and thus be useful for topological quantum computing \cite{fnote.126}. There are even experiments on InAs-superconductor systems whose results appear to be consistent with the existence of Majorana fermions \cite{Das.Nat.2012,Finck.PRL.2013}.

Thanks to the increased interest and activity, there have been numerous recent reports \cite{Shabani.PRB.2016,Shojaei.APL.2015,Tschirky.PRB.2017,Hatke.arXiv.2017} of improved quality of InAs 2DES samples, achieving low-temperature electron mobilities exceeding $2\times10^{6}$ cm$^{2}$/Vs \cite{Tschirky.PRB.2017}. Many of these reports demonstrate the high quality through the observation of well-developed integer quantum Hall effect (IQHE) states at high Landau level filling factors ($\nu$). Conspicuously absent, however, is any signature of the fractional quantum Hall effect (FQHE), a hallmark of interacting electrons. Here we report the observation of FQHE at $\nu=4/3$ in a recently fabricated, high-quality InAs 2DES \cite{Tschirky.PRB.2017}. Our observation places InAs among a handful of semiconductors which exhibit the FQHE: GaAs \cite{Tsui.PRL.1982,Pan.PRL.2002}, Si \cite{Nelson.APL.1992,Kott.PRB.2014,Lu.PRB.1970}, AlAs \cite{Lay.PRB.1993,DePoortere.APL.2002,Shayegan.PSS.2006,Chung.PRM.2017}, GaN \cite{Manfra.JAP.2002}, CdTe \cite{Piot.PRB.2010}, ZnO \cite{Tsukazaki.Nat.2010}, and Ge \cite{Shi.PRB.2015}; another 2DES hosting FQHE is, of course, graphene \cite{Du.Nat.2009,Bolotin.Nat.2009,Feldman.PRL.2013}. Besides being of general importance, the presence of FQHE in InAs 2DESs paves the way for the realization of \textit{parafermions} \cite{Lindner.PRX.2012, Clarke.Nat.2013, Mong.PRX.2014,Alicea.ARCMP.2015}, quasi-particles that are even more exotic than Majoranas. Prafermions, which can be viewed as fractionalized Majoranas, are excitations of \textit{interacting} topological systems such as FQHE edge states and, unlike the standard Majoranas, can be used to implement \textit{universal} topological quantum computation.


The details of our sample fabrication and parameters are reported in Ref. \cite{Tschirky.PRB.2017}. As shown in Fig. \ref{fig:fullfield} right inset, the sample hosts a 2DES in a 24-nm-wide InAs quantum well, grown on a nearly lattice-matched GaSb substrate via molecular beam epitaxy. Our sample is from wafer M as described in Table III of Ref. \cite{Tschirky.PRB.2017}. The structure starts with a 600-nm-thick GaSb layer, followed by a 10-period Al$_{0.33}$Ga$_{0.67}$Sb superlattice. The quantum well, located 25 nm below the surface, is flanked by a 200-nm-thick Al$_{0.33}$Ga$_{0.67}$Sb lower barrier and a 20-nm-thick AlSb upper barrier. The structure ends with a 5-nm-thick GaSb cap layer. The $8\times8$ \textbf{\textit{k}}$\cdot$\textbf{\textit{p}} simulations show that the structure is a quasi-single-interface quantum well \cite{Tschirky.PRB.2017}. The 2DES has density $n=7.8 \times 10^{11}$ cm$^{-2}$ and low-temperature mobility of $1.8 \times 10^6$ cm$^2$/Vs \cite{Tschirky.PRB.2017}. The sample we measured has a 4 mm$\times$ 4 mm van der Pauw geometry. Contacts to the 2DES are made by annealing In at 200 \textdegree C for 5 minutes. The transport measurements were carried out in a superconducting-resistive hybrid magnet system with a maximum field of 45 T, and a dilution refrigerator with a base temperature of $\simeq35$ mK. We used a low-frequency lock-in technique and an excitation current of $\sim100$ nA to measure the transport properties. 

\begin{figure}[b!]
  \begin{center}
    \psfig{file=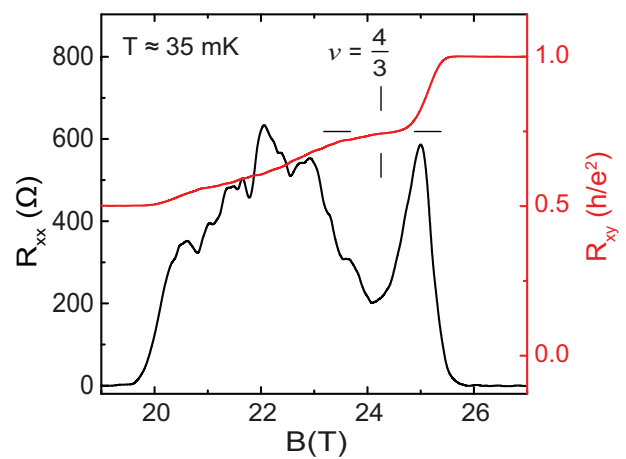, width=0.46\textwidth}
  \end{center}
  \caption{\label{fig2} $R_{xx}$ and $R_{xy}$ traces are shown near $\nu=4/3$. The vertical lines mark the expected magnetic field position for $\nu=4/3$. The horizontal lines indicate the expected corresponding $R_{xy}$ quantized value ($3h/4e^2$).}
  \label{fig:fraction}
\end{figure}

Figure \ref{fig:fullfield} highlights our main result: the observation of FQHE in an InAs 2DES. It shows the full-field longitudinal ($R_{xx}$) and Hall ($R_{xy}$) magnetoresistance traces up to 45 T at $T\simeq35$ mK. The vertical marks in Fig. \ref{fig:fullfield} indicate a number of filling factor positions in magnetic field. These marks are determined from the slope of $R_{xy}$ vs. $B$ at low magnetic fields, and they match all the resolvable $R_{xx}$ minima positions to within 1$\%$. Between 19 and 26 T, as shown in expanded plots in Fig. \ref{fig:fraction}, we observe a deep $R_{xx}$ minimum at $\nu=4/3$, signaling a developing FQHE at this filling. This minimum is concomitant with a nearly quantized $R_{xy}$ plateau with a Hall resistance within $1\%$ of the expected value (3$h/4e^2$). 

The FQHE at $\nu=4/3$ is observed thanks to the very high quality of the sample, as demonstrated by the left inset in Fig. \ref{fig:fullfield} which captures more details of the data at lower magnetic fields. The inset includes $R_{xx}$ and $R_{xy}$ traces taken at $n=8.1 \times 10^{11}$ cm$^{-2}$ and $T\simeq70$ mK; these were measured in a different system with a superconducting magnet only. They show that the Shubnikov-de Haas oscillations are resolvable at magnetic fields as low as $\simeq0.5$ T ($\nu=68$). Also, the oscillations are spin-resolved down to $B\simeq0.9$ T ($\nu=37$). These features attest to the very high quality of the 2DES.

\begin{figure}[t!]
  \begin{center}
    \psfig{file=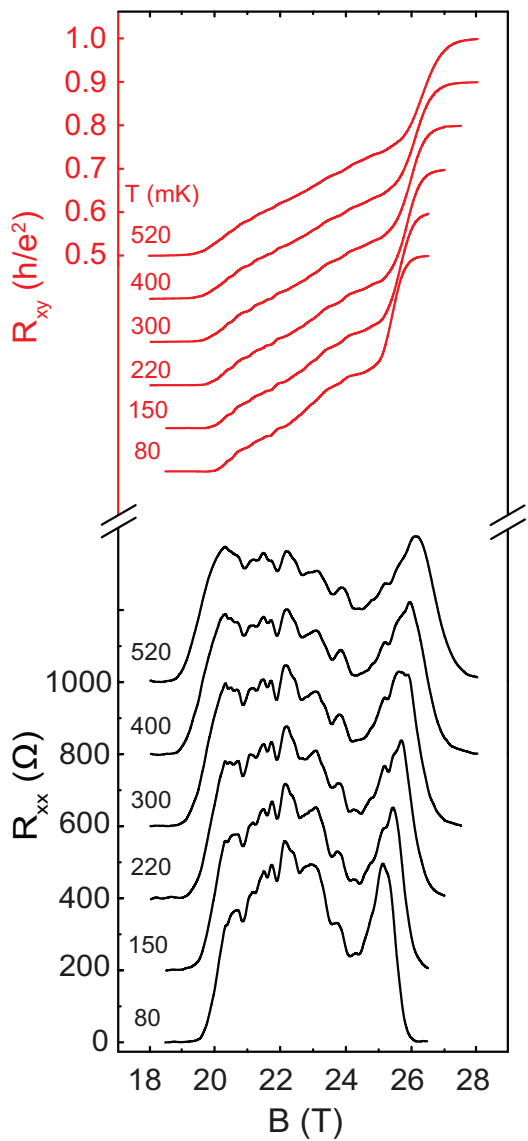, width=0.4\textwidth}
  \end{center}
  \caption{\label{fig3} Temperature dependence for the $R_{xx}$ and $R_{xy}$ traces between $\nu=1$ and $\nu=2$. For clarity, starting from the 80 mK trace, each $R_{xx}$ trace is shifted up vertically by 0.2 k$\Omega$. Similarly, each $R_{xy}$ trace is shifted vertically by $0.1(h/e^2)$ as well.}
  \label{fig:temp}
\end{figure}

\begin{figure}[b!]
  \begin{center}
    \psfig{file=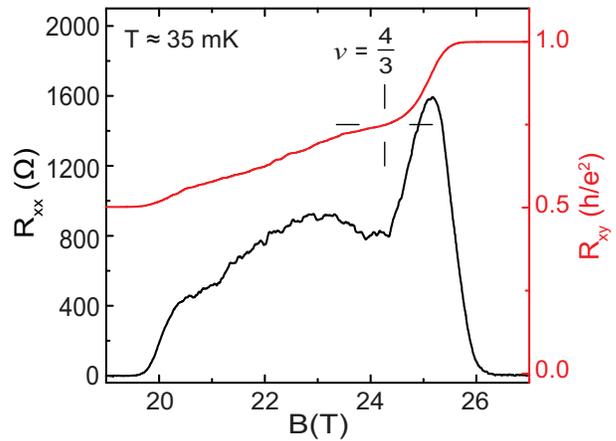, width=0.46\textwidth}
  \end{center}
  \caption{\label{fig4} $R_{xx}$ and $R_{xy}$ traces are shown near $\nu=4/3$ for a different cool-down and a different current-voltage configuration compared to Fig. \ref{fig:temp}. The vertical lines mark the expected magnetic field position for $\nu=4/3$, and the horizontal lines indicate the expected quantized $R_{xy}$ value ($3h/4e^2$).}
  \label{fig:additional}
\end{figure}

Aside from the pronounced low-field oscillations, however, we also observe wide $R_{xx}$ minima and $R_{xy}$ plateaus in Fig. 1 at intermediate and high magnetic fields. At the highest fields, above 26 T, there is an extremely wide $R_{xx}$ minimum, accompanied by a $\nu=1$ quantized Hall plateau, spanning over a field range of $\simeq20$ T and likely even beyond. These very wide plateaus indicate that the Landau levels are significantly broadened and that, despite the very high mobility, there is still a fair amount of disorder present. Assuming the remote, ionized background impurities and defects \cite{Furukawa.APL.1994,Shen.APL.1995} in the barriers to be the main scattering source \cite{Tschirky.PRB.2017}, a rough estimate based on experimental data and calculations for mobility vs. density in GaAs 2DESs \cite{Shayegan.APL.1988,Jiang.APL.1988,Pfeiffer.APL.1989,Umansky.JCG.2009,Gardner.JCG.2016, Hwang.PRB.2008} gives an ionized impurity/defect concentration in the barriers of $\simeq5\times10^{16}$ cm$^{-3}$ for the InAs sample in our present study. For comparison, the best GaAs samples grown in ultra-clean molecular beam epitaxy chambers have a background impurity concentration of $\simeq1\times10^{13}$ cm$^{-3}$, nearly four orders of magnitude smaller \cite{Shayegan.APL.1988,Jiang.APL.1988,Pfeiffer.APL.1989,Umansky.JCG.2009,Gardner.JCG.2016, Hwang.PRB.2008}. This still relatively high amount of disorder in InAs samples gives a tentative explanation for the weakness of the $\nu=4/3$ FQHE and the absence of FQHE at more filling factors. We also examined another sample with a similar structure but narrower well-width (15 nm), with density $n=8.6 \times 10^{11}$ cm$^{-2}$ and mobility $0.98 \times 10^6$ cm$^2$/Vs. No features except for a single $R_{xx}$ maximum is observed between $\nu=1$ and $\nu=2$. 

The $R_{xx}$ and $R_{xy}$ data in Fig. 1 exhibit two other noteworthy features. First, we observe an asymmetry in the well-quantized, wide Hall plateaus at intermediate and high magnetic fields, namely, the plateaus extend farther on the high-field side of a given integer filling factor. Similar asymmetries have been reported for 2DESs in other materials with moderate quality \cite{Wei.PRL.1988,Koch.PRB.1991,Goldammer.JCG.1999}. We note that, as might be expected, the plateaus we observe in Fig. 1 data are in fact reasonably symmetric in \textit{filling factor} with respect to integer values. For example, the $\nu=2$ Hall plateau extends from $\nu=2.37$ ($B=13.6$ T) to $\nu=1.64$ ($B=19.7$ T); these $\nu$ are nearly equidistant from $\nu=2$ ($B=16.1$ T). Regardless of its origin, the plateau asymmetry can explain the absence of a $\nu=5/3$ FQHE in Figs. \ref{fig:fullfield} and \ref{fig:fraction} despite the presence of a clear $\nu=4/3$ FQHE: The field position for $\nu=5/3$ ($B=19.4$ T) overlaps with the extended $\nu=2$ plateau \cite{fnote.50}.  

Another noteworthy feature in the data of Figs. 1 and 2 is the presence of numerous small, oscillatory features seen in $R_{xx}$ in the field range $20<B<25$ T. As best seen in the temperature dependence data shown in Fig. \ref{fig:temp}, these features are reproducible, although their amplitude diminishes at higher temperatures. This reproducibility implies that the oscillations are not random noise. In our experiments we find that, while the oscillations are reproducible in a given cool-down from room temperature and for a fixed current-voltage configuration, their detailed features (field positions and amplitudes) depend on the cool-down and the measurement configuration. Figure \ref{fig:additional} shows $R_{xx}$ and $R_{xy}$ traces for a different cool-down and with a different current-voltage configuration. The small oscillatory features have changed but the $R_{xx}$ minimum and $R_{xy}$ quantization near $\nu=4/3$ are consistently reproducible. A more thorough study is needed to discern the origin of these intriguing oscillations. 

Finally, we highlight the temperature evolution of the $R_{xx}$ minimum at $\nu=4/3$. Figure 3 data indicate that, as temperature decreases, the $R_{xx}$ minimum only becomes relatively deeper, meaning that the resistance value at the minimum remains nearly constant while the resistance maxima on both flanks increase. A qualitatively similar behavior is seen for other FQHE states when they are not particularly strong \cite{Willett.PRL.1987}. Unfortunately, this temperature dependence precludes us from measuring an accurate gap energy for the $\nu=4/3$ FQHE.  

In summary, we report the observation of FQHE in an InAs 2DES, namely a deep $R_{xx}$ minimum at $\nu=4/3$ along with a nearly quantized Hall plateau at the expected value ($3h/4e^2$). This observation is an important step towards the realization of exotic quasi-particles such as parafermions which could find use in the realization of universal topological quantum computing. The relative fragility of the observed $\nu=4/3$ FQHE, and the absence of other FQHE states, however, highlight the significant disorder still present in the currently available samples and the need for future improvements.

\begin{acknowledgments}

We acknowledge support from the NSF Grants ECCS 1508925, DMR 1305691, and DMR 1709076. Our research was also funded in part by QuantEmX grants from ICAM and the Gordon and Betty Moore Foundation through Grant GBMF5305 to M. K. M., Md. S. H., and M. S. Our measurements were partly performed at the National High Magnetic Field Laboratory (NHMFL), which is supported by the NSF Cooperative Agreement DMR 1157490, by the State of Florida, and by the DOE. We thank T. Murphy, J. H. Park, H. Baek and G. E. Jones at NHMFL for technical assistance, and Javad Shabani for illuminating discussions. 

\end{acknowledgments}


\begin{thebibliography}{99}
\bibitem{fnote.50} As seen most clearly in Fig. \ref{fig:fullfield}, there is a faint shoulder near $3h/5e^2$ in $R_{xy}$ around 21.9 T in magnetic field. However, this $R_{xy}$ feature is not at the correct magnetic field position for $\nu=5/3$ ($B=19.4$ T); nor is it accompanied by an $R_{xx}$ minimum.
\end{thebibliography}
\end{document}